\documentclass[11pt]{article}
\usepackage{moriond,epsfig}

\def\Journal#1#2#3#4{{#1} {\bf #2}, #3 (#4)}

\def\NPB{{\em Nucl. Phys.} B}

\def\PRL{\em Phys. Rev. Lett.}
\def\PRD{{\em Phys. Rev.} D}

\def\be{\begin{equation}}
\def\ee{\end{equation}}
\def\bea{\begin{eqnarray}}
\def\eea{\end{eqnarray}}

\begin{document}
\vspace*{4cm}
\title{ROLE OF NONPERTURBATIVE INPUT IN QCD RESUMMED HEAVY BOSON $Q_T$
DISTRIBUTION}

\author{JIANWEI QIU$^a$ and XIAOFEI ZHANG$^b$}

\address{$^a$Department of Physics and Astronomy,
             Iowa State University\\
             Ames, Iowa 50011, U.S.A.\\
         $^b$Center for Nuclear Research, 
             Department of Physics,
             Kent State University\\
             Kent, Ohio 44242, U.S.A.}

\maketitle
\abstracts{
We show that role of nonperturbative input in the 
$b$-space QCD resummation formalism for heavy boson transverse
momentum ($Q_T$) distribution strongly depends on 
collision energy $\sqrt{S}$.  At collider energies,
the larger $\sqrt{S}$ is, the weaker role nonperturbative input plays,
and better predictive power the $b$-space resummation formalism has. 
}
\section{Introduction}

With new data coming from Fermilab Run II and from the Large Hadron
Collider (LHC) in the near future, we expect to test 
Quantum Chromodynamics (QCD) to a new level of accuracy, 
and also expect that a better understanding of QCD 
will underpin precision tests of the Electroweak interactions 
and particle searches beyond the Standard Model.  
In this talk, we will concentrate on Drell-Yan type production of 
color neutral heavy boson ($W^\pm$, $Z$, and Higgs) of invariant mass
$Q$ at small transverse momentum $Q_T$.

When $Q_T \ll Q$, the $Q_T$ distribution of the heavy boson 
production calculated in the conventional fixed-order
perturbation theory receives a large logarithm, 
$\ln(Q^2/Q_T^2)$.  Beyond the leading order, we can get two 
powers of the logarithm for every power of $\alpha_s$.
Therefore, at sufficiently small $Q_T$, convergence of the
conventional perturbative expansion in powers of $\alpha_s$ is
impaired, and the logarithms must be resummed.\cite{QCD-rpt}   

\section{The $b$-space resummation formalism}

By using the renormalization group equation technique, 
Collins, Soper, and  Sterman (CSS) derived a $b$-space
resummation formalism for the $Q_T$ distribution of the
heavy boson production.\cite{CSS-Resum}
The formalism has the following generic form 
for collisions between hadrons $A$ and $B$, 
\begin{equation}
\frac{d\sigma_{A+B\rightarrow V+X}}{dQ^2\, dy\, dQ_T^2} 
= 
\frac{d\sigma_{A+B\rightarrow V+X}^{\rm (resum)}}{dQ^2\, dy\, dQ_T^2} 
+
\frac{d\sigma_{A+B\rightarrow V+X}^{\rm (Y)}}{dQ^2\, dy\, dQ_T^2}\, , 
\label{css-gen}
\end{equation}
where $V$ represents the heavy boson.\cite{CSS-Resum}  
In Eq.~(\ref{css-gen}),
the $\sigma^{\rm (Y)}$ term is negligible for small $Q_T$ and becomes
important when $Q_T\sim Q$.  The $\sigma^{\rm (resum)}$ includes all
orders resummation of the large logarithms
and can be expressed as \cite{CSS-Resum}
\be
\frac{d\sigma_{A+B\rightarrow C+X}^{\rm (resum)}}{dQ^2\, dy\, dQ_T^2} 
=
\frac{1}{(2\pi)^2}\int d^2b\, e^{i\vec{Q}_T\cdot \vec{b}}\,W(b,Q)
= \int \frac{db}{2\pi}\, J_0(Q_T\, b)\, bW(b,Q)
\label{css-resum}
\ee
where $J_0$ is Bessel function and 
$W(b,Q)=\sum_{ij}\sigma_{ij\rightarrow V}^{(0)}(Q)\, W_{ij}(b,Q)$
is a $b$-space distribution with dependence on rapidity $y$
suppressed.  The $\sigma_{ij\rightarrow V}^{(0)}(Q)$ is the 
lowest order partonic cross section for partons of flavor $i$ and $j$
to produce a heavy boson $V$ of invariant mass $Q$. 

Because of initial-state hadrons, $W_{ij}(b,Q)$ depends on momentum
scale of hadron wave function (1/fm~$\sim \Lambda_{\rm QCD}$) and is 
in principle nonperturbative.  However, when $b$ is small 
($\ll 1/\Lambda_{\rm QCD}$), physics associated with 
momentum scales $1/b$ and $Q$ are perturbative, and large logarithms 
from $\log(1/b^2)$ to $\log(Q^2)$ can be resummed by solving the 
following evolution equation \cite{CSS-Resum}
\be
\frac{\partial}{\partial \ln Q^2}\, W_{ij}(b,Q) 
=
\left[ K(b\mu,\alpha_s)+G(Q/\mu,\alpha_s) \right]\ W_{ij}(b,Q)
\ee
where kernels $K$ and $G$ themselves obey renormalization group
equations.\cite{CSS-Resum} By solving the linear evolution equation,  
one derives the resummed $b$-space distribution, 
$
W(b,Q) = {\rm e}^{-S(b,Q)}\, W(b,c/b),
$
with constant $c={\cal O}(1)$ and $S(b,Q) = \int_{c^2/b^2}^{Q^2}\, 
  \frac{d\bar{\mu}^2}{\bar{\mu}^2} \left[
  \ln\left(\frac{Q^2}{\bar{\mu}^2}\right) 
     A(\alpha_s(\bar{\mu})) + B(\alpha_s(\bar{\mu})) \right]$.
The $A(\alpha_s(\bar{\mu}))$ and $B(\alpha_s(\bar{\mu}))$ 
are perturbatively calculable in power series of $\alpha_s$. 
All large logarithms in $W(b,Q)$ are completely resummed
into the exponential factor $\exp[-S(b,Q)]$ leaving $W(b,c/b)$
with only one hard scale $1/b$.  
When $b\leq b_{max}\ll 1/\Lambda_{\rm QCD}$, the 
nonperturbative physics in $W(b,c/b)$ can be factorized into parton
distributions, and the resummed $W(b,Q)$ can be factorized
as \cite{CSS-Resum} 
\be
W^{\rm pert}(b,Q)
=
\sum_{ij} \sigma_{ij\rightarrow V}^{(0)}
\left[f_{a/A}\otimes C_{a\rightarrow i}\right]
\otimes
\left[f_{b/B} \otimes C_{b\rightarrow j} \right]
\times {\rm e}^{-S(b,Q)}
\label{css-pert}
\ee
where $f$ and $C$ are parton distributions and perturbatively 
calculable coefficient functions, respectively.  In
Eq.~(\ref{css-pert}), the $\otimes$ represents convolution over parton
momentum fraction, and the superscript ``{\rm pert}'' 
indicates that $W^{\rm pert}(b,Q)$ is perturbatively calculable
at small $b$ if parton distributions are known. 

\begin{figure}
\hskip 0.2in
\begin{minipage}[t]{3.2in}
  \psfig{figure=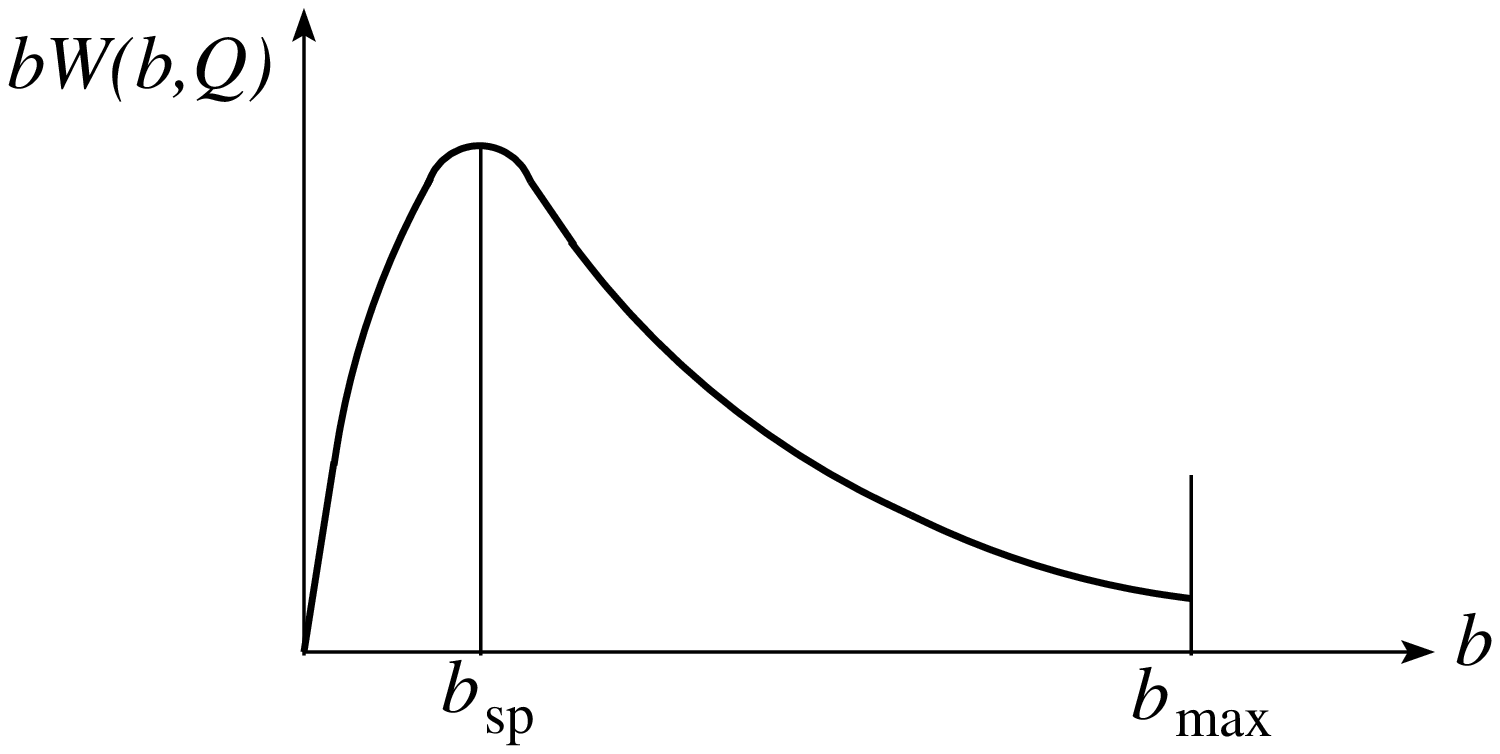,width=3.2in}
\vspace{-0.25in}
\caption{Generic resummed $b$-space distribution.} 
\label{fig1}
\end{minipage}
\hspace*{0.2in}
\begin{minipage}[t]{2.4in}
\hskip 0.1in
  \psfig{figure=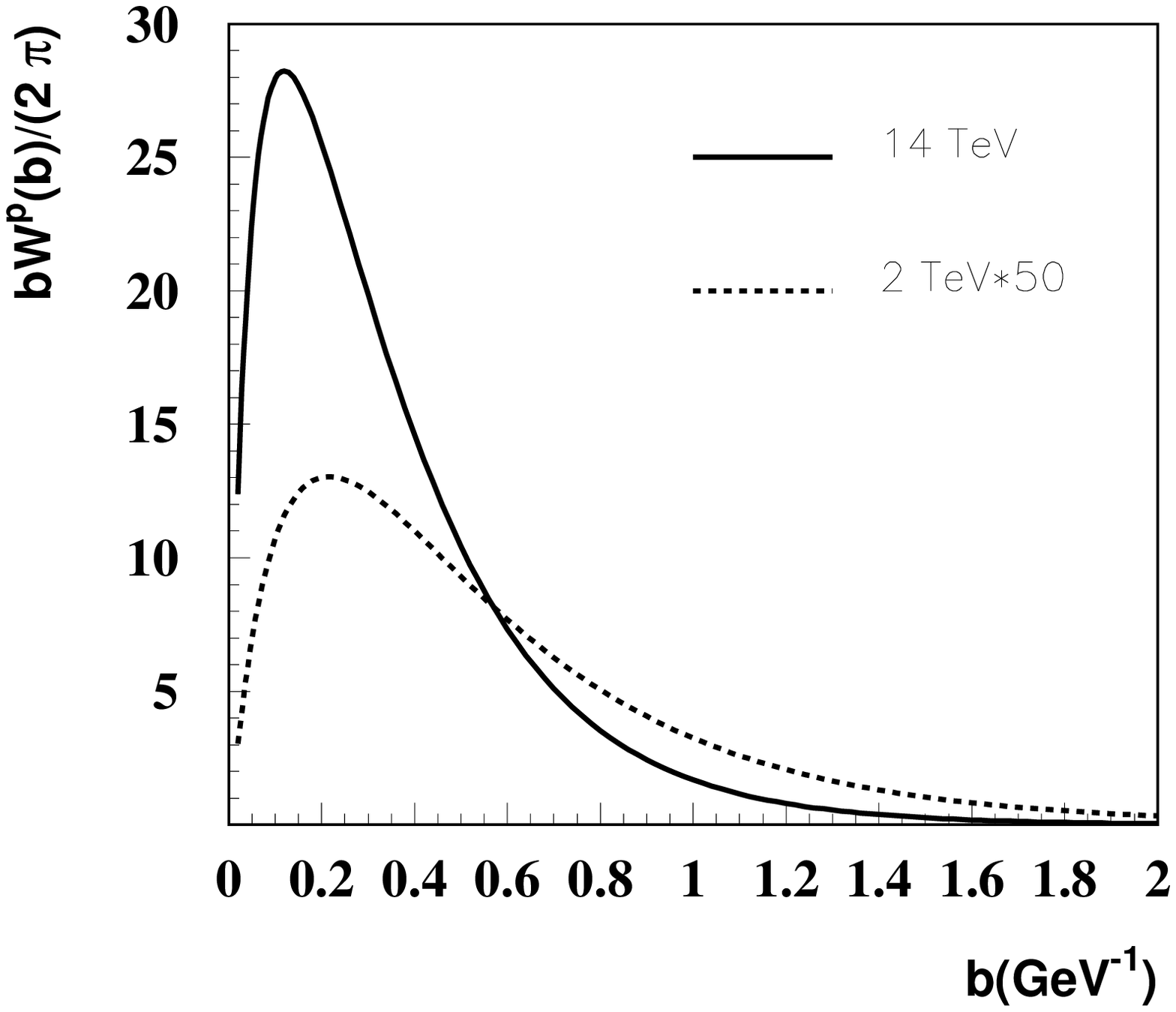,width=2.2in}
\vspace{-0.18in}
\caption{Resummed $b$-space distribution at Tevatron and LHC
energies.} 
\label{fig2}
\end{minipage}
\hfill
\vspace{-0.13in}
\end{figure}

When $Q^2$ is large enough, the perturbatively resummed $b$-space
distribution $W^{\rm pert}(b,Q)$ has a generic functional form shown
in Fig.~\ref{fig1}.  The peak and corresponding saddle point 
($b_{\rm sp}$) depends on values of $Q$ and $\sqrt{S}$.\cite{QZ-Resum}
Since the $W^{\rm pert}(b,Q)$ is only reliable for small $b$ region,
an extrapolation to large $b$ is necessary in order to
complete the Fourier transform in Eq.~(\ref{css-gen}).

\section{Extrapolation to large $b$ region}

Collins, Soper, and Sterman proposed the following large-$b$
extrapolation \cite{CSS-Resum}
\begin{equation}
W^{\rm CSS}(b,Q) \equiv W(b_*,Q)\,F^{NP}(b,Q)\, ,
\label{css-W-b}
\end{equation}
where $b_*\equiv b/\sqrt{1+(b/b_{max})^2} < b_{max} \sim
0.5$~GeV$^{-1}$ and $F^{NP}(b,Q)\sim\exp(-\kappa b^2)$ is a
Gaussian-like nonperturbative function.  The $\kappa$ depends on 
fitting parameters, $g_i$ with $i=1,..,n$.  
By adjusting functional form for $\kappa$ and fitting parameters $g_i$,
$Q_T$ distributions derived from $W^{\rm CSS}(b,Q)$ are not
inconsistent with Fermilab data on $Z$ and $W^\pm$.\cite{LBLY}

Although it is successful in interpreting existing data, 
the $b$-space resummation formalism has been questioned due to  
two apparent drawbacks.\cite{QCD-rpt}  
The first is the difficulty of matching
the resummed and fixed-order predictions; and the second is to know
the quantitative difference between the prediction and the fitting
because of the introduction of a nonperturbative $F^{NP}$.  
Recently, we demonstrated that both apparent drawbacks can be  
overcome.\cite{QZ-Resum}  

According to the large-$b$ extrapolation 
defined in Eq.~(\ref{css-W-b}), the 
nonperturbative function $F^{NP}$ and its fitting parameters can not
only affect the large $b$ region, but also significantly change the
perturbatively calculated $b$-space distribution at small
$b$.\cite{QZ-Resum} 
In order to quantitatively separate QCD prediction from 
parameter fitting, we introduce a new large-$b$
extrapolation \cite{QZ-Resum}  
\be
W(b,Q) = \left\{
\begin{array}{ll}
 W^{\rm pert}(b,Q) &  \mbox{$b\leq b_{max}$} \\
 W^{\rm pert}(b_{max},Q)\,
 F^{NP}(b,Q;b_{max})
                      &  \mbox{$b > b_{max}$}.
\end{array} \right. 
\label{qz-W-b}
\ee
This new extrapolation preserves the QCD resummed $b$-space
distribution at small $b$.  For large $b$ region, a new functional
form of $F^{NP}$ was derived by adding power corrections to the
evolution equations of $W(b,Q)$ \cite{QZ-Resum}
\be
F^{NP}=
\exp\left\{
 -\ln(\frac{Q^2 b_{max}^2}{c^2}) \left[
   g_1 \left( (b^2)^\alpha - (b_{max}^2)^\alpha\right)
  +g_2 \left( b^2 - b_{max}^2\right) \right] 
 -\bar{g}_2 \left( b^2 - b_{max}^2\right)
\right\} .
\label{qz-fnp}
\ee
The $(b^2)^\alpha$ term with $\alpha<1/2$ corresponds to a direct
extrapolation of resummed $W^{\rm pert}(b,Q)$, 
while the $b^2$ terms correspond to the
power corrections to the evolution equation.  The $g_2$ term
corresponds to power correction from 
soft gluon shower; and the $\bar{g}_2$ is due to parent partons'
nonvanish intrinsic transverse momentum.  The parameters, $g_1$ and
$\alpha$ are completely fixed by $W^{\rm pert}$ by requiring the
first and second derivatives of $W(b,Q)$ to be continuous at
$b=b_{max}$.  

\section{Predictive power of the formalism}

In order to exam the predictive power, we divide the
$b$-integration in Eq.~(\ref{css-resum}) into a perturbative ($b\leq
b_{max}$) and a nonperturbative ($b>b_{max}$) region.  
Predictive power of the $b$-space resummation formalism 
is sensitive to the relative contributions from these two regions. 

From the generic $b$-space distribution in Fig.~\ref{fig1}, better
predictive power requires a smaller $b_{\rm sp}$ for the saddle
point.  We found that numerical value of $b_{\rm sp}$ has a strong 
dependence on the $\sqrt{S}$ in addition to its
well-known $Q^2$ dependence.\cite{QZ-Resum}  The larger
$\sqrt{S}$ corresponds to much a smaller $b_{\rm sp}$.  In
Fig.~\ref{fig2}, we plot the $b$-space distribution for $Z$ production
at two different collision energies, $\sqrt{S}=14$~TeV (solid) and 
$\sqrt{S}=2.0$~TeV (dashed) with $W(b,Q)$ at Tevatron energy
multiplied by a factor of 50.  The $W(b,Q)$ at LHC energy is clearly
peaked at a smaller $b_{\rm sp}$.

\begin{figure}
\hskip 0.1in
\begin{minipage}[t]{2.9in}
\hskip 0.2in \psfig{figure=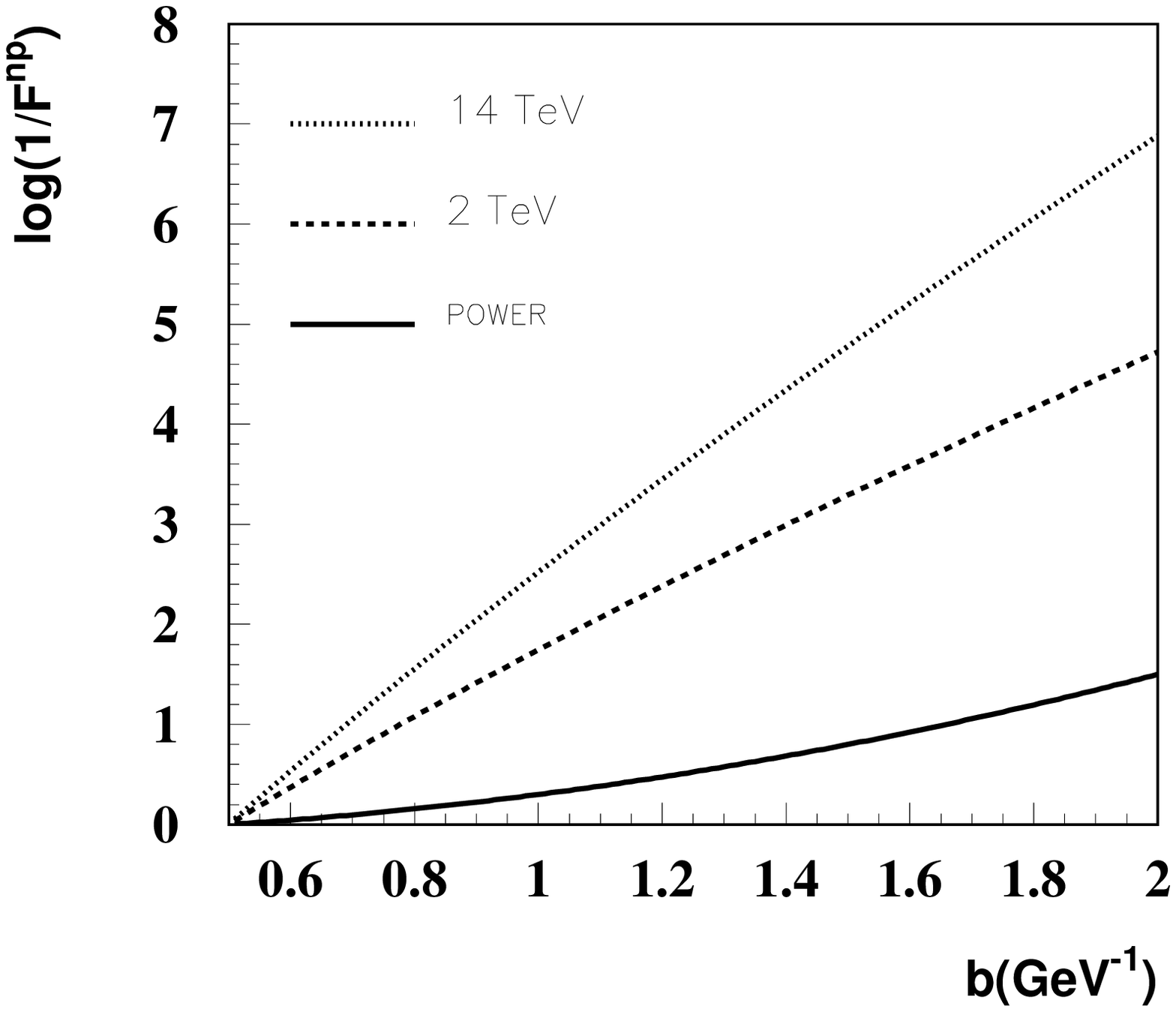,width=2.4in}
\vspace{-0.12in}
\caption{Nonperturbative $\log(1/F^{NP})$ defined in 
         Eq.~(\protect\ref{qz-fnp}) as a function of $b$.}
\label{fig3}
\end{minipage}
\hspace*{0.2in}
\begin{minipage}[t]{3.0in}
\hskip 0.1in
   \epsfig{figure=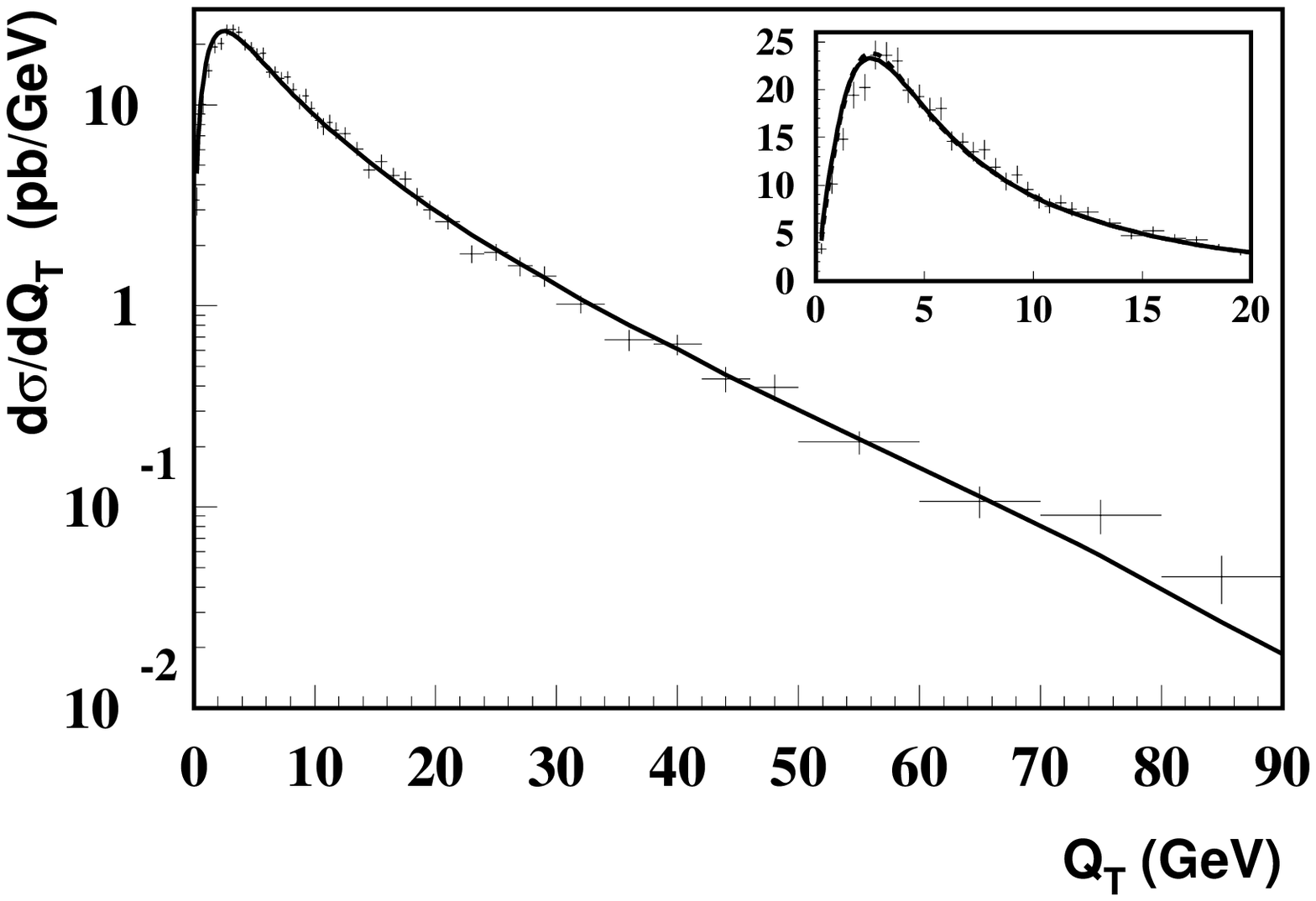,width=2.8in}
\vspace{-0.1in}
\caption{Comparison between resummed $Q_T$ distribution and CDF data
         on $Z$.} 
\label{fig4}
\end{minipage}
\hfill
\vskip -0.2in
\end{figure}

Precise contribution from large $b$ region depends on the functional
form and corresponding parameters of the $F^{NP}$.  
In Fig.~\ref{fig3}, we plot $\log(1/F^{NP})$ as a
function of $b$ for $Q=M_Z$ and $b_{max}=0.5$~GeV$^{-1}$.  The dotted
(dashed) line represents the leading fractional power term at
$\sqrt{S}=14$~TeV ($\sqrt{S}=2$~TeV).  Both power correction terms are
combined into the solid line.  The parameters, $g_2$ and
$\bar{g}_2$, are fixed by fitting low energy Drell-Yan
data.\cite{QZ-Resum} 
Since the $b$-integration converges at $b\leq 
2$~GeV$^{-1}$ for all $Q_T<Q$, we expect very small
power corrections for heavy boson production at
collider energies, in particular, at the LHC energy.

Because of its weak role in $F^{NP}$, 
we can first neglect the power corrections and
predict the heavy boson transverse momentum distribution without any
free parameter, except the choice of $b_{max}$.  Variation of
$b_{max}$ is a good test of uncertainties of our predictions.  
In Fig.~\ref{fig4}, we compare our prediction with Fermilab data on
$Z$ production with $b_{max}=0.5$~GeV$^{-1}$ (solid line).  We find
that the theoretical prediction is insensitive to the choice of
$b_{max}$ within 0.3 to 0.8; and the power corrections in $F^{NP}$
only change the $Q_T$ distribution in Fig.~\ref{fig4} for less than
5\% at the lowest $Q_T$ and less than 1\% for
$Q_T>5$~GeV.\cite{QZ-Resum}   As expected from the features
shown in Figs.~\ref{fig2} and \ref{fig3}, the power corrections to 
$Z$ production at the LHC at $\sqrt{S}=14$~TeV is less than 1\% even
at the lowest $Q_T$ bin.\cite{ZG-Resum}

\section{Conclusions}

We conclude that CSS $b$-space resummation formalism 
with a new large-$b$ extrapolation has an excellent predictive 
power for heavy boson transverse momentum distribution at collider
energies.  Larger the collision energy is, better the
predictive power is.  At collider energies, the large-$b$
nonperturbative contribution is dominated by the
extrapolation of $W^{\rm pert}(b,Q)$, and power corrections plays a
very weak role.

\section*{Acknowledgments}
This work was supported in part by the U.S. Department of Energy under
Grant Nos. DE-FG02-86ER40251 and DE-FG02-87ER40371.



\section*{References}
\begin{thebibliography}{99}

\bibitem{QCD-rpt} 
S. Catani, {\it et al.}, in the Report of the ``1999 CERN Workshop on
SM Physics (and more) at the LHC'', hep-ph/0005025, and references
therein. 

\bibitem{CSS-Resum}
J.C. Collins, D.E. Soper, and G. Sterman, 
\Journal{\NPB}{250}{199}{1985}.

\bibitem{QZ-Resum}
J.-W. Qiu and X.-F. Zhang, 
\Journal{\PRL}{86}{2724}{2001};
\Journal{\PRD}{63}{114011}{2001}, and references therein.

\bibitem{LBLY}
F. Landry, R. Brock, G. Ladinsky, and C.-P. Yuan,
\Journal{\PRD}{63}{013004}{2001}.

\bibitem{ZG-Resum}
X.-F. Zhang and G. Fai, 
{\em Phys. Rev. C}\ (in press), hep-ph/0202029. 

\end{thebibliography}
\end{document}